\documentclass[aps,prl,preprint,superscriptaddress,showpacs]{revtex4-1}
\usepackage[greek,english]{babel}
\usepackage{multirow}
\usepackage{graphicx}
\usepackage{amsmath}
\usepackage{slashed}
\newcommand{\helacnlo}{\texttt{HELAC-NLO}}

\newcommand{\pythia}{\texttt{PYTHIA}}
\newcommand{\herwig}{\texttt{HERWIG}}
\newcommand{\lhapdf}{\texttt{LHAPDF}}

\newcommand{\powhel}{\texttt{PowHel}}
\newcommand{\powhegbox}{\texttt{POWHEG-BOX}}
\newcommand{\fastjet}{\texttt{FastJet}}

\newcommand{\mev}{\ensuremath{\,\mathrm{MeV}}}
\newcommand{\gev}{\ensuremath{\,\mathrm{GeV}}}
\newcommand{\tev}{\ensuremath{\,\mathrm{TeV}}}

\newcommand{\pT}{\ensuremath{p_{\perp}}}

\newcommand{\pTj}{\ensuremath{p_{\perp}^{j}}}

\newcommand{\pTmiss}{\ensuremath{\slash\hspace*{-5pt}{p}_{\perp}}}

\newcommand{\mt}{\ensuremath{m_{\rm t}}}

\newcommand{\bt}{\ensuremath{\bar{{\rm t}}}}
\newcommand{\bb}{\ensuremath{\bar{{\rm b}}}}

\newcommand{\ttA}{t$\bar{{\rm t}}\gamma$}
\newcommand{\ttZ}{t$\bar{{\rm t}}Z$}
\newcommand{\ttH}{t$\bar{{\rm t}}H$}
\newcommand{\ttj}{t$\bar{{\rm t}}$+jet}

\newcommand{\NLO}{{\rm NLO}}
\newcommand\Ref[1]     {Ref.\,\cite{#1}}
\newcommand\Refs[1]    {Refs.\,\cite{#1}}

\newcommand\fig[1]     {Fig.\,{\ref{#1}}}

\def\beq{\begin{equation}}
\def\eeq{\end{equation}}
\def\bsp#1\esp{\begin{split}#1\end{split}}
\def\bal#1\eal{\begin{align}#1\end{align}}

\begin{document}

\preprint{LPN 11-91}

\title{$Z^0$-boson production in association with a
top anti-top pair at NLO accuracy with parton shower effects}
\date{\today}
\author{M.~V.~Garzelli}
\affiliation{Institute of Physics, University of Debrecen, H-4010 Debrecen P.O.Box 105, Hungary}
\affiliation{Laboratory for Astroparticle Physics, University of Nova Gorica,
SI-5000 Nova Gorica, Slovenia}
\author{A.~Kardos}
\affiliation{Institute of Physics, University of Debrecen, H-4010 Debrecen P.O.Box 105, Hungary}
\affiliation{Institute of Nuclear Research of the Hungarian Academy of Sciences, Hungary}
\author{C.~G.~Papadopoulos}
\affiliation{Institute of Nuclear Physics, NCSR 
Demokritos
         GR-15310 Athens, Greece}
\author{Z.~Tr\'ocs\'anyi}
\affiliation{Institute of Physics, University of Debrecen, H-4010 Debrecen P.O.Box 105, Hungary}
\affiliation{Institute of Nuclear Research of the Hungarian Academy of Sciences, Hungary}

\begin{abstract}
We present predictions for the production cross section of 
a Standard Model $Z^0$-boson in association with a
t\bt\ pair at the next-to-leading order accuracy in QCD, matched with shower 
Monte Carlo programs to evolve the system down to the hadronization 
energy scale.
We adopt a framework based on three well established numerical
codes, namely the \powhegbox, used for computing the cross section,
\helacnlo, which generates all necessary input matrix elements, and finally a
parton shower program, such as \pythia\ or \herwig, which 
allows for including t-quark and $Z^0$-boson decays 
at the leading order accuracy and generates shower emissions,
hadronization and hadron decays.
\end{abstract}


\maketitle

With increasing collider energies, the t-quark plays an increasingly
important role in particle physics. Its production cross section grows
faster with energy than that of any other discovered 
Standard Model (SM) particle.
Already after the first year of successful run of the LHC, the t\bt\ 
production 
cross section is measured with unprecedented accuracy at $\sqrt{s} = 7$\,TeV, 
so that the corresponding SM theoretical prediction will be challenged soon 
\cite{Saleem:2011fu,Chatrchyan:2011yy}.
However, many other t-quark properties have not yet been directly accessed.
In particular, its couplings to neutral gauge (especially the $Z^0$)
and scalar bosons  
are still prone to large uncertainties.  
In \Refs{Baur:2004uw,Baur:2005wi} the possibility of
measuring the \ttZ\ and \ttA\ couplings was studied based upon
leading-order (LO) parton level predictions. Although such precision is
sufficient for feasibility studies, finding the optimal values of the
experimental cuts requires indeed predictions at higher accuracy.

An essential step towards higher accuracy is the inclusion of 
next-to-leading order (NLO) radiative corrections. Recent theoretical advances
made possible our computation of the $pp \to$
\ttZ\ 
cross section at the parton level, including 
QCD corrections at NLO \cite{Kardos:2011na}. 
In order though to get the
optimum benefit and to produce predictions that can be directly
compared to experimental data at the hadron level, a matching with
parton shower (PS) and hadronization implemented in shower Monte Carlo
(SMC) programs is ultimately inevitable. Thus, in this letter we
present first predictions for $pp \to$ \ttZ\ production at LHC
at the matched NLO + PS accuracy.
 
In constructing a general interface of PS to matrix element (ME)
computations with NLO accuracy in QCD, we have chosen to combine the POWHEG
\cite{Nason:2004rx,Frixione:2007vw} method and FKS subtraction scheme \cite{Frixione:1995ms}, as implemented in
the \powhegbox\ \cite{Alioli:2010xd} computer framework, 
with the HELAC-NLO~\cite{Bevilacqua:2010mx}
approach, respectively.  
In particular, \powhegbox\ requires
the relevant MEs as external input. We obtain the latter in
a semi-automatic way by codes in the \helacnlo\ package \cite{Bevilacqua:2011xh}.              
With this input \powhegbox\ is used to generate events 
at the Born plus first radiation emission level, stored in Les Houches Event
Files (LHEF) \cite{Alwall:2006yp}, that can be
interfaced to standard 
SMC programs. Previous applications of the whole framework, 
proving
its robustness, were presented in \Refs{Kardos:2011qa,Garzelli:2011vp}.
This same setup also allows for exact NLO pure hard-scattering 
predictions.  
Further details on the implementation of the computation of the
pp $\rightarrow$ \ttZ\ hard-scattering cross-section in it, 
at NLO accuracy in QCD, 
together with checks, were recorded in \Ref{Kardos:2011na}.
\begin{figure*}
\begin{center}
$
\begin{array}{cc}
  \includegraphics[width=8.4cm]{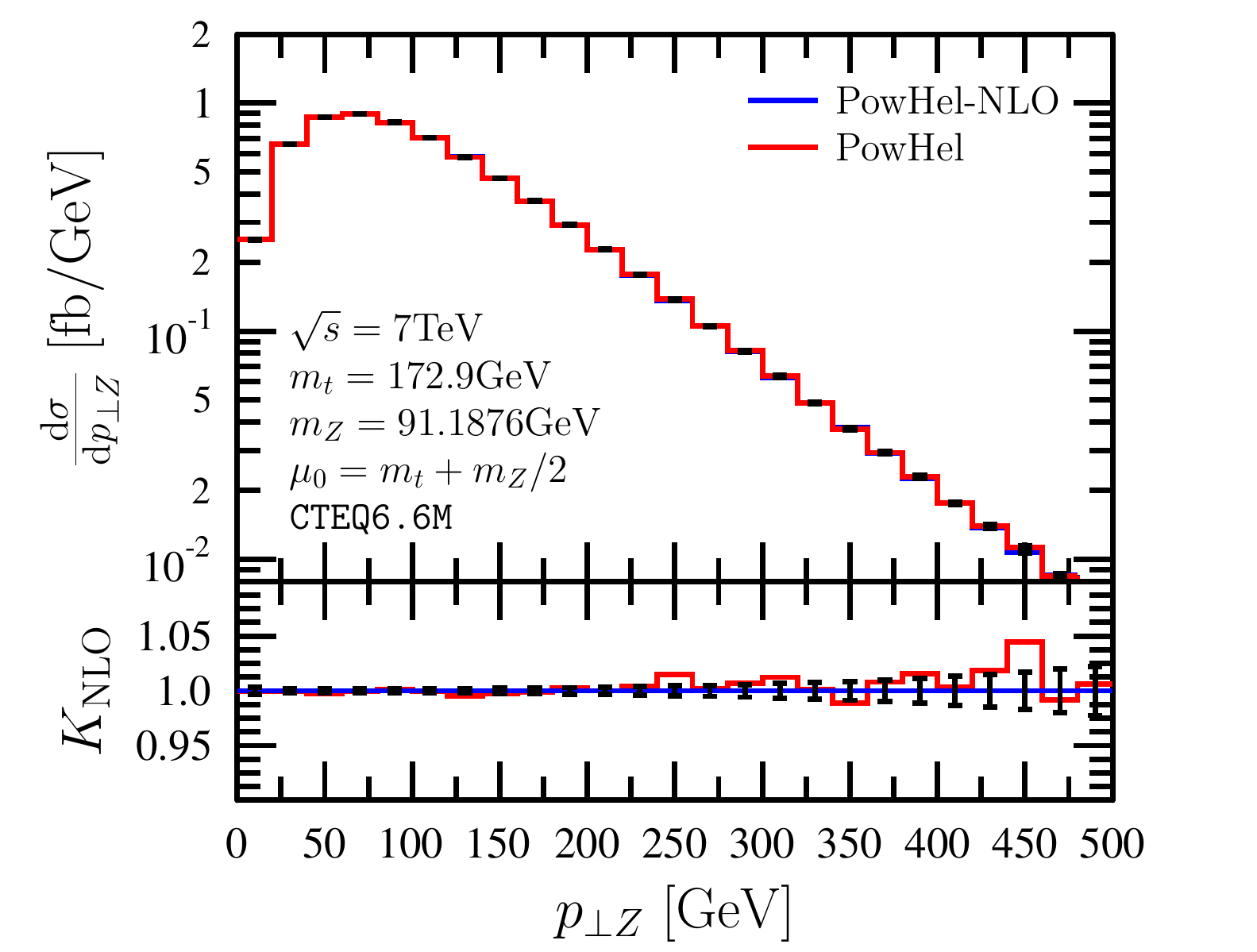}
& \includegraphics[width=8.4cm]{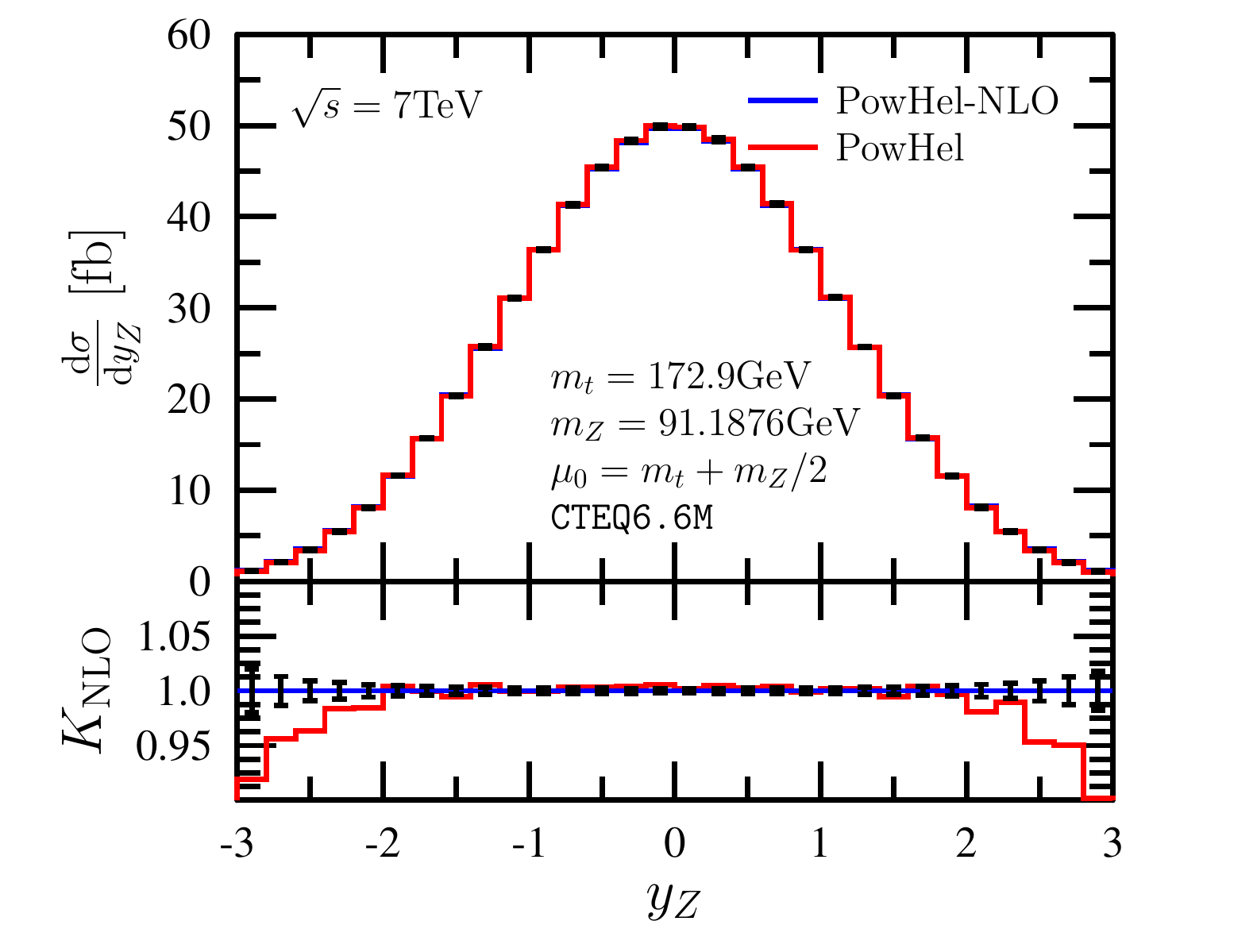}
\end{array}
$
\caption{Transverse momentum (left) and rapidity (right) distributions
of the $Z^0$-boson at NLO and after first radiation (PowHel).
The lower panels show the ratio of the two predictions with combined
uncertainties.}
\label{fig:LHE-NLOcomp}
\end{center}
\end{figure*}

\newpage
All these computations are steps of an ongoing 
project for generating event samples for $pp \to$ t\bt$X$ processes, \
where $X$ stays for a hard partonic object. 
The events we generate are stored in LHEF, made available
on the web, and are ready to be interfaced to standard SMC programs to 
produce predictions for distributions at the hadron level. 
Such predictions can be useful for optimizing the selection cuts applied 
to disentangle the signal from the background, in order 
to improve the experimental accuracy of the t-quark
coupling measurements.

Interfacing NLO calculations to SMC programs allows to estimate
the effects of decays, shower emissions and hadronization, therefore we
have analyzed the process at hand at three different stages of evolution:
\begin{itemize}
\item[]
{\bf PowHel:} we analyzed the events including no more parton
emissions than the first and hardest one,
collected in LHEF produced as output of \powhegbox + \helacnlo\ (\powhel).
\item[]
{\bf Decay:} we just included on-shell decays of t-quarks
and the $Z^0$-boson, as implemented in \pythia~\cite{Sjostrand:2006za}, 
and further decays
of their decay products, like charged leptons (the $\tau$ is considered  
as unstable) and gauge bosons ($W$), turning off any initial and final PS
and hadronization effect.
\item[]
{\bf Full SMC:} decays, shower evolution, hadronization and hadron
decays have been included in our simulations, using both \pythia\ and 
\herwig~\cite{Corcella:2002jc}.
\end{itemize}

In our computation, we adopted the following parameters:
$\sqrt{s} = 7\tev$, \texttt{CTEQ6.6M} PDF set from \lhapdf{}, with a
2-loop running $\alpha_s$, 5 light flavours and
$\Lambda_5^{\overline{\mathrm{MS}}} = 226\mev$, $m_t = 172.9\gev$,
$m_Z = 91.1876\gev$, $G_F = 1.16639\cdot 10^{-5}\gev^{-2}$. The
renormalization and factorization scales were chosen equal to the default
scale $\mu_0 = m_t + m_Z/2$.  We used the last version of the SMC fortran
codes: \pythia\ 6.425 and \herwig\ 6.520. 
Following our implementation of \ttH\ hadroproduction 
in \Ref{Garzelli:2011vp}, in both SMC setup muons (default in \pythia) and neutral pions were assumed as stable particles. All other particles and hadrons were
allowed to be stable or to decay according to the default
implementation of each SMC. Masses and total decay widths of the
elementary particles were tuned to the same values in \pythia\ and
\herwig, but each of the two codes was allowed to compute autonomously
partial branching fractions in different decay channels for all
unstable particles and hadrons. Multiparticle interaction effects were
neglected (default in \herwig). Additionally, the intrinsic
\pT-spreading of valence partons in incoming hadrons in \herwig\ was
assumed to be 2.5~GeV. 

First, to check event generation, we compared several distributions from events including no more           than 
first radiation emission (\powhel\ level) with
the \NLO\ predictions of \Ref{Kardos:2011na}. We found agreement for
all considered distributions. As examples, we show in \fig{fig:LHE-NLOcomp} the
transverse momentum and rapidity distributions of the $Z^0$-boson. 

\begin{figure*}
\begin{center}
$
\begin{array}{cc}
  \includegraphics[width=8.4cm]{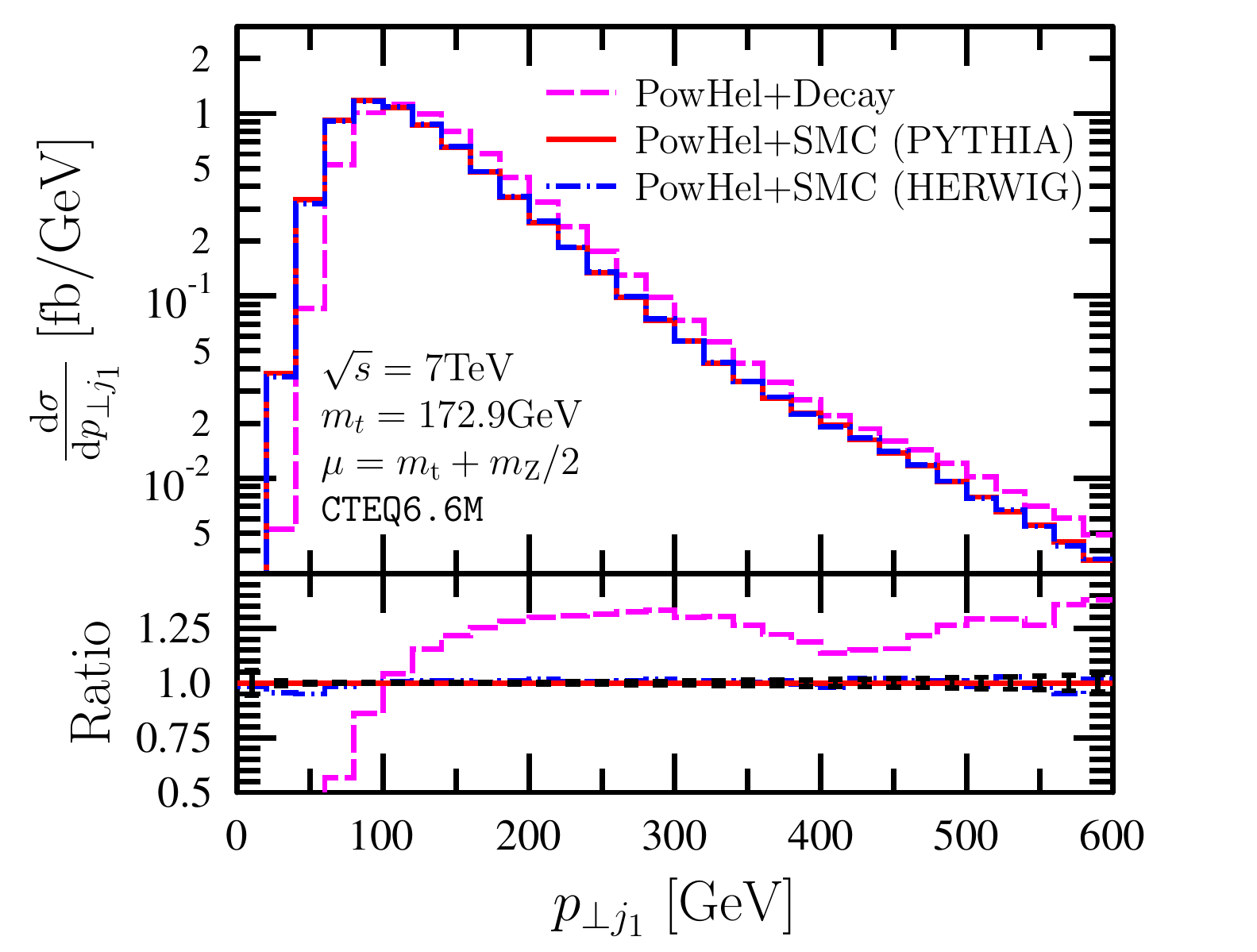}
& \includegraphics[width=8.4cm]{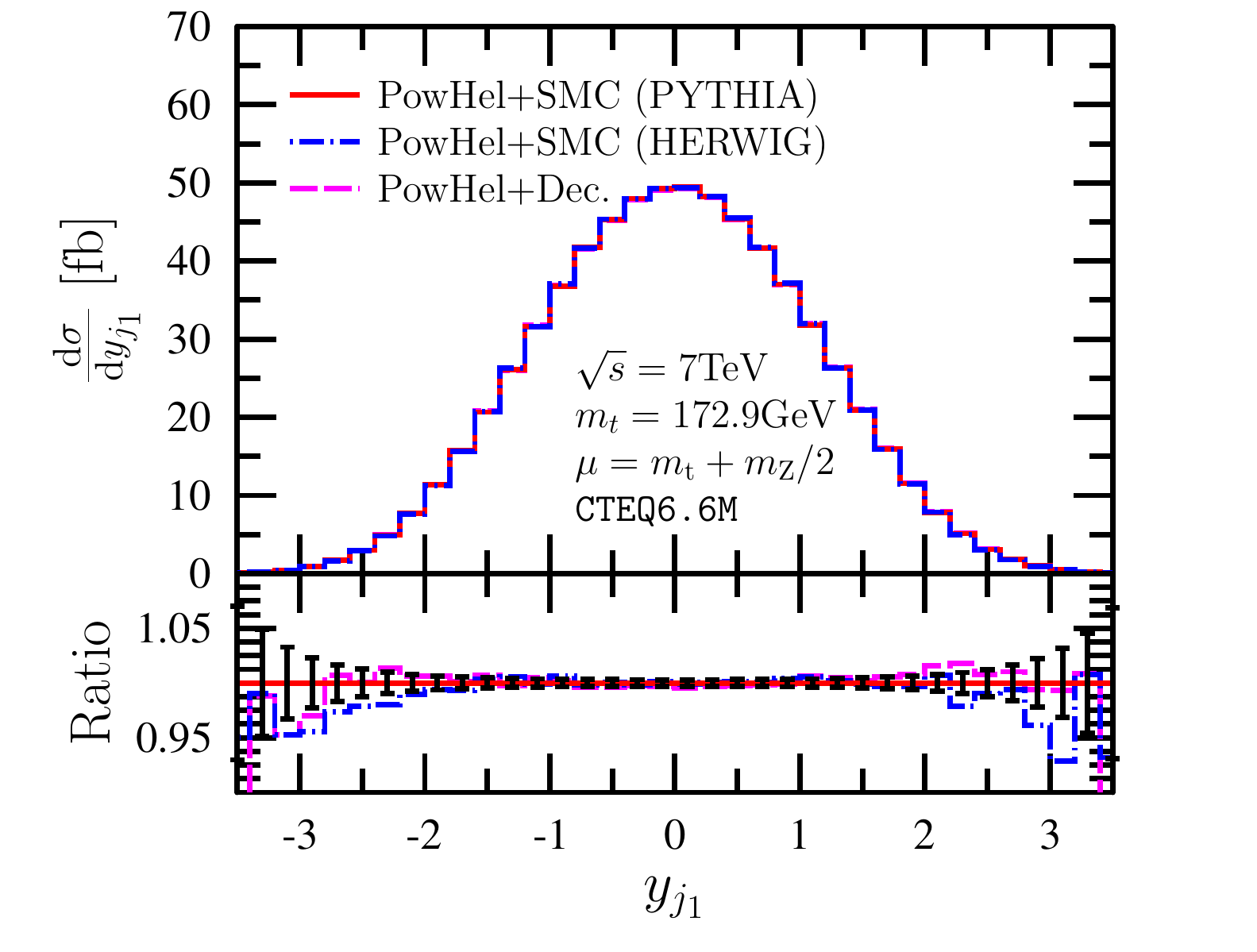}
\end{array}
$
\caption{Transverse momentum (left) and rapidity (right) distributions 
of the hardest jet after
decay and after full SMC. The lower panels show the ratio of all
predictions to \powhel+SMC using \pythia.}
\label{fig:pTj}
\end{center}
\end{figure*}

Next, we studied the effect of the full SMC by comparing distributions
at the decay and SMC level. Since particle yields are very
different at the end of these two stages, we made such a comparison
without any selection cut, in order to avoid the introduction of any
bias. As an illustrative example, we present the distributions of the
transverse momentum and rapidity of the hardest jet, \pTj\ and $y_j$,
in \fig{fig:pTj}. 
Jets are reconstructed through the $\mathrm{anti}- k_\bot$
algorithm with $R=0.4$, as implemented in \fastjet\
\cite{Cacciari:2008gp}. The softening of the transverse
momentum spectrum is apparent as going from the decay level to the full SMC one, while the effect of the shower on the rapidity of the hardest jet is almost
negligible and rather homogeneous. The cross-section at both level 
amounts to $\sigma = $  138.7 $\pm$  $0.01$\,fb.
Using our setup for the full SMC's, we found agreement between \pythia\
and \herwig\ predictions within very few percent, 
despite the conceptual differences between 
the two SMC generators as for the shower ordering variables 
and hadronization models, confirming the level of
agreement already reported in \Ref{Garzelli:2011vp} in the study of 
a different process. 

Next, we made predictions for \ttZ\ hadroproduction
at the LHC including experimental selection cuts.
For this analysis,
in the absence of a dedicated tune for NLO 
matched computations, \pythia\ was tuned to the Perugia 2011 set of values, 
one of the most recent LO tunes~\cite{Skands:2010ak}, updated on the basis 
of recent LHC data, providing a \pT-ordered PS.
Its application turned out to increase 
our particle yields by about 10\,\%. As a consequence, the agreement 
between the tuned \pythia\ and untuned \herwig\ predictions decreases
(as for \herwig, the default configuration was used, providing instead an
angular-ordered PS), and we present only the \pythia\ ones.

In case of \ttZ\ hadroproduction overwhelming backgrounds come from
t\bt +jets final states.  In \Ref{Baur:2005wi} the
differential cross section as a function of missing transverse momentum
for the production of $\pTmiss b \bar{b}$+4 jets was found a useful tool for
differentiating the signal and the possible backgrounds. The proposed
set of selection cuts is rather exclusive and the rates decrease 
so much 
that the measurement for the present LHC 
run
at $\sqrt{s}
= 7\tev$ 
looks quite demanding from the statistical point of view, 
therefore, we restrict ourself to present predictions for the
future runs at $\sqrt{s}= 14\tev$ ($\sigma_{\rm{all\,cut},14}$/$\sigma_{\rm{all\,cut},7} \sim$ 7 and 8 at the decay and at the full SMC level, respectively).  

\begin{figure*}
\begin{center}
$
\begin{array}{cc}
\includegraphics[width=8.4cm]{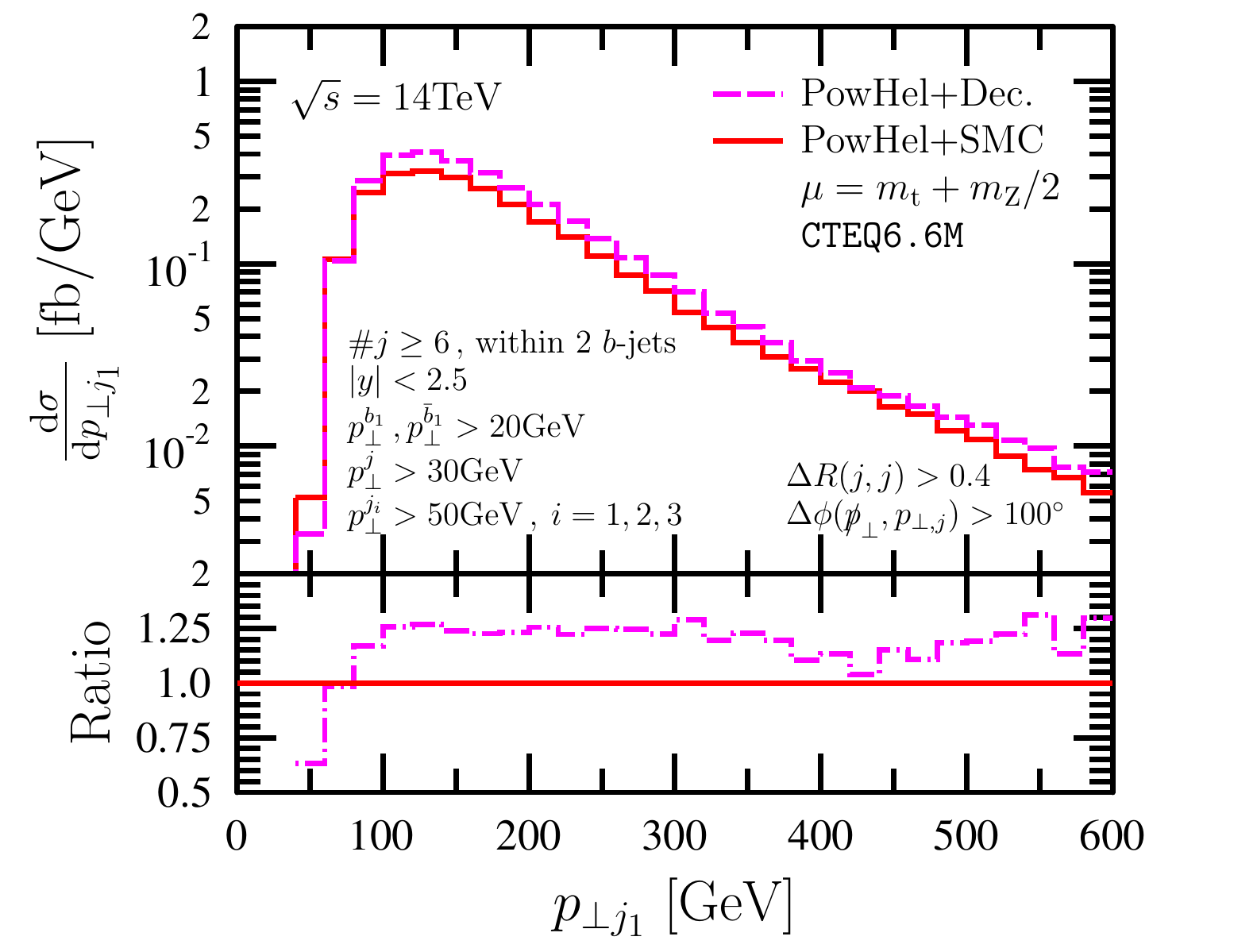}
&
\includegraphics[width=8.4cm]{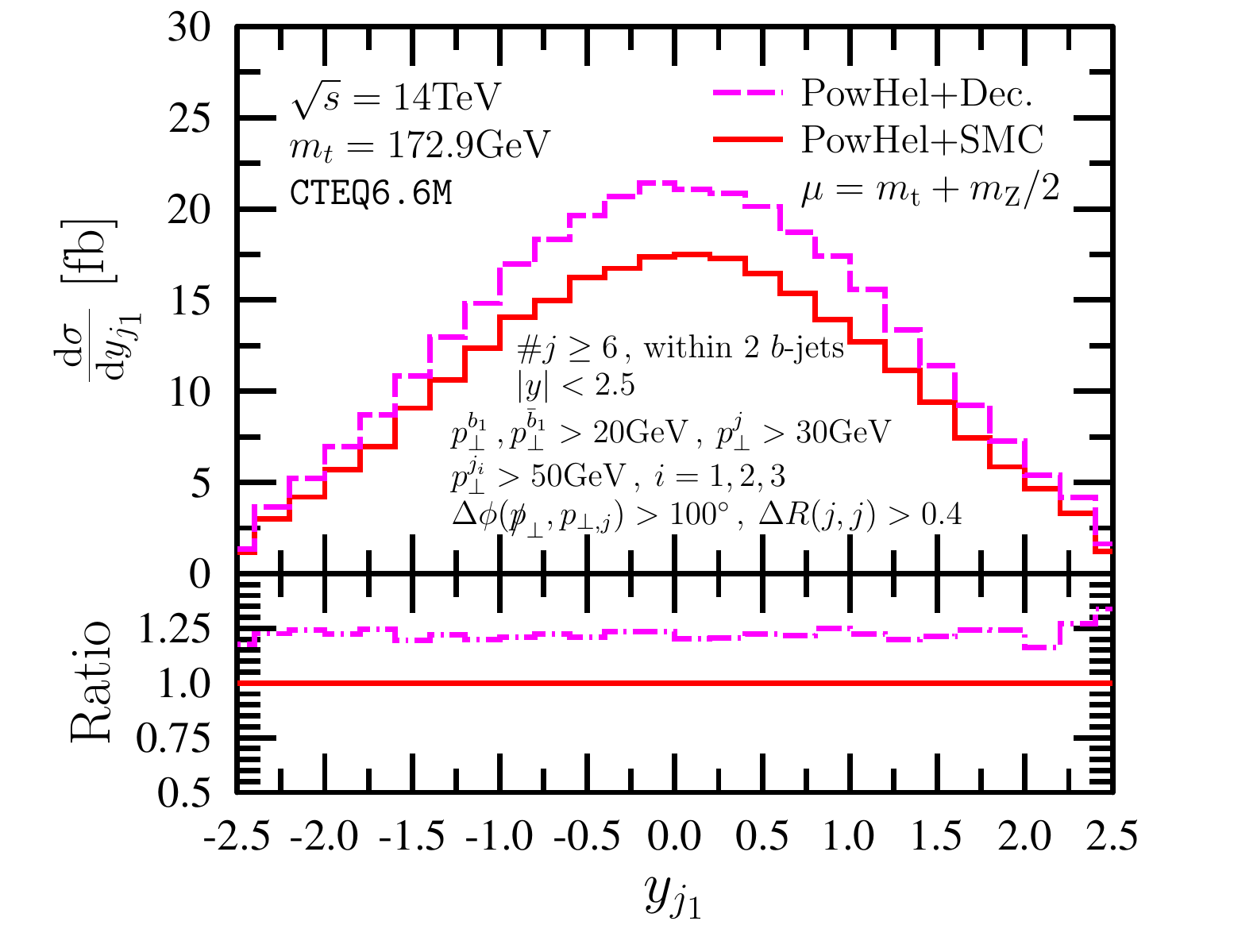}
\end{array}
$
\caption{Transverse momentum (left) and rapidity (right) distributions 
of the hardest jet after decay and after full SMC (\pythia), under 
selection cuts (1--8) implemented at both levels. The lower panels show 
the ratio of the predictions at different~levels.}
\label{fig:ptj1wcuts}
\end{center}
\end{figure*}

In \fig{fig:ptj1wcuts}
we 
show the
distributions of transverse momentum and rapidity of the hardest jet
using the following reduced set of cuts:
1) we reconstruct at least six jets with ra\-pi\-di\-ty $|y| < 2.5$,
2) of these we require at least one $b$-jet and one $\bar{b}$-jet,
3) for $b$-jets $\pT^b > 20$\gev,
4) for other jets $\pT^{{\rm non}\,\, b} > 30$\gev,
5) at least 3 jets ($b$ or non-$b$) with $\pTj > 50$\gev,
6) 
$\Delta R(j,j) > 0.4$, where $j$
denotes any ($b$ or non-$b$) jet and $\Delta R$ is defined as
$\sqrt{{\Delta\phi}^2+{\Delta y}^2}$,
7~--~8) $\Delta\phi(\slashed{p}_\bot,p_{\bot,j})> 100^\circ$, with
$p_{\bot,j}$ meaning either ($p_\bot (\hat{b}_1) + p_\bot (\hat{\bar{b}}_2)$)
(cut 7), or
($p_\bot (\hat{j}_1) + p_\bot (\hat{j}_2) + p_\bot (\hat{j}_3) + p_\bot (\hat{j}_4) $)
(cut 8), 
where $\hat{b}_1$, $\hat{\bar{b}}_2$ and $\hat{j}_1$, 
$\hat{j}_2$, $\hat{j}_3$, $\hat{j}_4$ are the jets that
allow for the best t $\to \rm{b} W^+$ $\to bjj$  and
\bt $\to \bb W^-$ $\to \bar{b}jj$ invariant mass simultaneous 
reconstruction, since they~minimize~the 
\bal
\bsp
\chi^2(b_1j_1j_2;\bar{b}_2j_3j_4)& =
  \frac{(m_{j_1j_2}-m_W)^2}{\sigma_W^2}
+ \frac{(m_{j_3j_4}-m_W)^2}{\sigma_W^2}
\\ &
+ \frac{(m_{b_1j_1j_2}-m_W)^2}{\sigma_{\rm t}^2}
+ \frac{(m_{\bar{b}_2j_3j_4}-m_W)^2}{\sigma_{\rm t}^2}
\esp
\nonumber
\eal
computed by considering all possible $j_kj_l$, $b_ij_kj_l$ and
$\bar{b}_ij_kj_l$ combinations.
The $W \to jj$ and t$ \to bjj$ invariant mass resolutions were set to
$\sigma_W = 7.8$\gev\ and $\sigma_t = 13.4$\gev\ , 
respectively~\cite{Beneke:2000hk}.
The \powhel+\pythia\ cross sections after these cuts amount to
$\sigma_{{\rm dec}} = 65.56 \pm 0.15$\,fb and
$\sigma_{{\rm SMC}} = 53.74 \pm 0.13$\,fb.

\begin{figure}
\begin{center}
\includegraphics[width=0.9\textwidth]{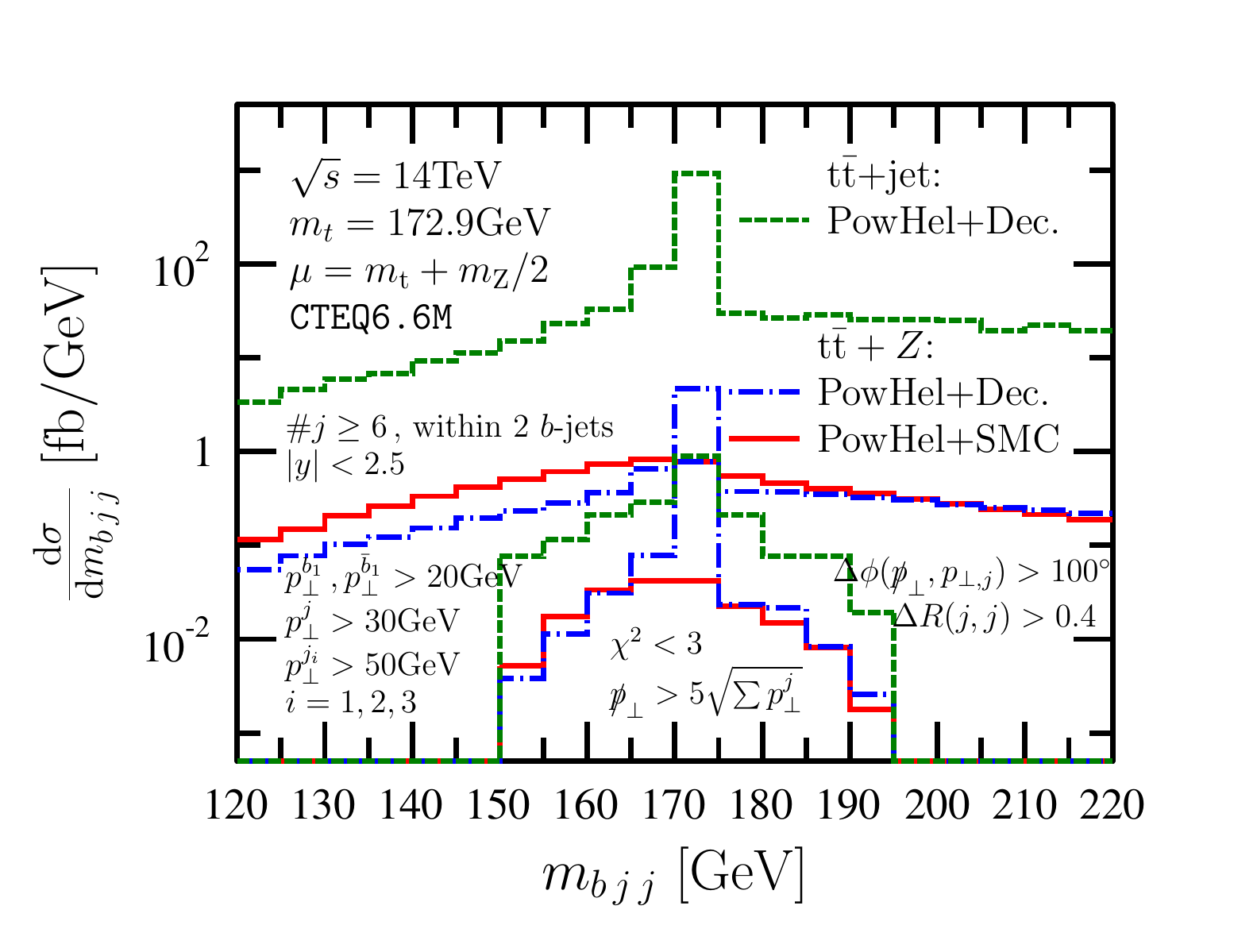}
\caption{Invariant mass distribution of the t-quark reconstructed from
the decay products at both decay and full SMC levels, 
for the \ttZ\ signal and, at the decay level,
 for one background (\ttj) after selection cuts (1--8)
(wider distributions in abscissa values) and after selection cuts (1--10)
(narrower distributions).}
\label{fig:mbjj}
\end{center}
\end{figure}

In \fig{fig:mbjj} we plot the invariant mass distribution of the
t-quark, as reconstructed from its decay products, 
by minimizing the $\chi^2$ above.
At the decay level, 
the reconstruction leads to a clear
peak centered around the \mt\ value. 
On the other hand, after full SMC, due both to further
      emissions which 
modify jet content and to
hadron decays, 
there are more candidate 
jets and the reconstruction is less successful. Although a peak is still 
visible (more evident in non-log scale), it is smeared towards lower mass 
values. 
The effect of the shower and hadronization 
turns out to be especially large in the peak region.

\begin{figure}
\begin{center}
\includegraphics[width=0.9\textwidth]{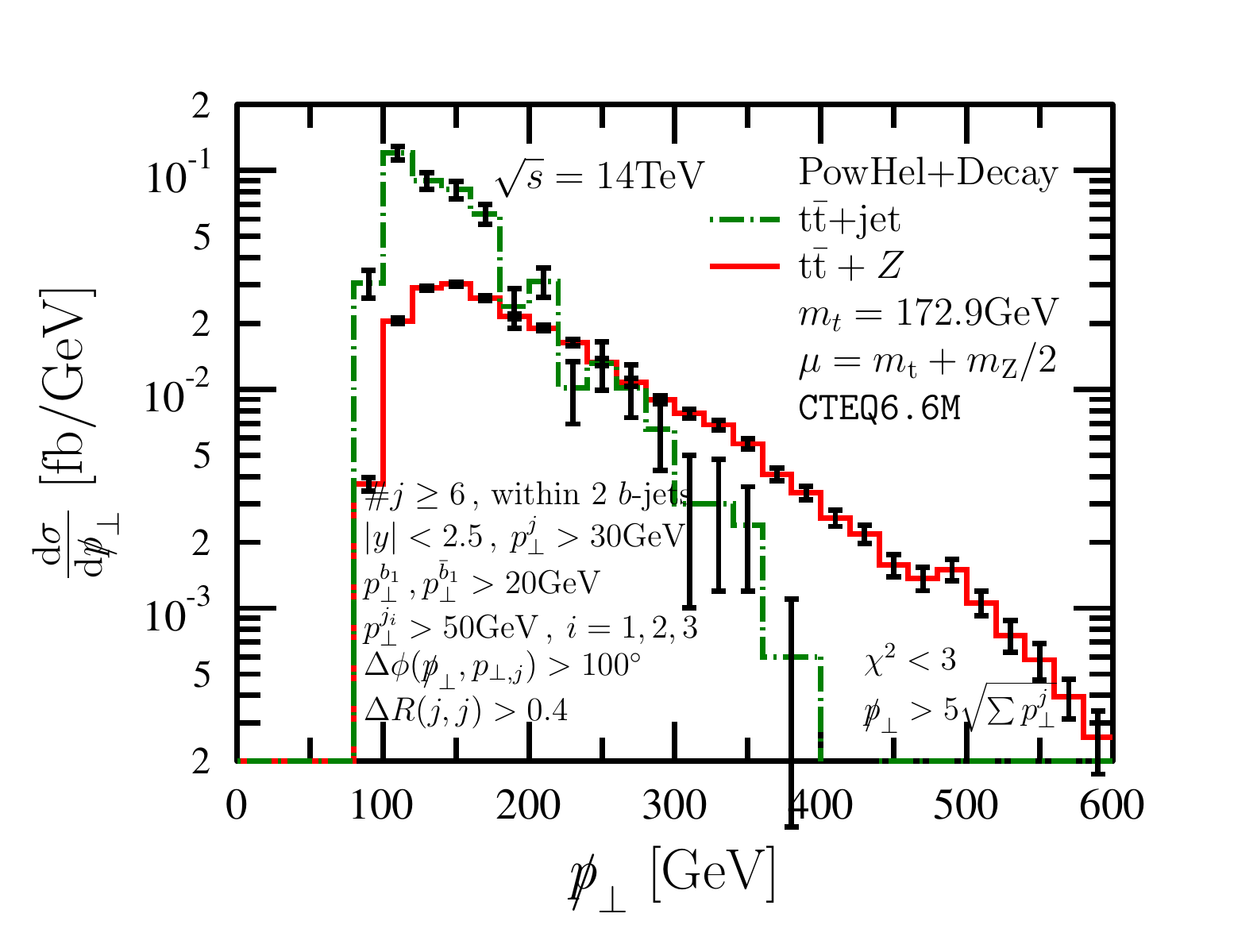}
\caption{Distribution of the missing transverse momentum after decay,
under physical cuts (1--10) 
applied to the signal (\ttZ) and to one background (\ttj).}
\label{fig:ptmiss}
\end{center}
\end{figure}

In \fig{fig:mbjj} we also show the $m_{bjj}$ distribution 
after decay for an important background process: 
t\bt-pair production associated with a jet 
(obtained at the scale $\mu_0 = m_t$).  Clearly, the
background overwhelms the signal, therefore, in order to select the
peak region, we include two more cuts:
9)        \pTmiss (due to
all $\nu$'s) $> 5 \sqrt{\sum \pTj}$ (of all jets, $b$ or
non-$b$), and
10) $\chi^2_{\min} < 3$, where $\chi^2_{\min}$ is the minimum of the
$\chi^2$ above.
Thus, we closely reproduce the cuts in \Ref{Baur:2005wi},
aimed at favoring
 the $Z^0 \to \nu\nu$ decay channel.
The
effect of the whole set of cuts on top reconstruction in \ttZ\ and 
\ttj\ events is also shown in \fig{fig:mbjj}. 
Although this set of cuts is
effective in selecting the signal, the background is globally
still larger:
for the signal
$\sigma_{{\rm dec}} = 4.83 \pm 0.04$\,fb,
while for the background
$\sigma_{{\rm dec}} = 9.86 \pm 1.05$\,fb.
However, as can be understood 
from \fig{fig:ptmiss}, where the distributions of the
missing transverse momentum after decay are shown
for both \ttZ\ and \ttj, these 
cuts allow for disentangling the signal, at least at the decay level. 
At the shower level, the $\pTmiss$ distributions of the \ttZ\ signal 
still shows a harder spectrum than the one of the \ttj\ background, but to
a lesser extent.
In this case, the effect of different top 
reconstruction strategies, 
still under investigation, 
can be~crucial~to 
help 
better disentangle the signal from the background in the 
$\pTmiss b\bar{b}$+4 jets considered channel. 

We studied the hadroproduction of a $Z^0$ boson in
association with a t\bt-pair, process of interest for
measuring the \ttZ-coupling directly at the LHC. We studied the effect
of heavy particle decays as well as the one of the full SMC.
We produced predictions for the LHC. As the production cross section is
rather small, measuring the \ttZ-coupling becomes more feasible 
after the planned 14\,TeV energy upgrade. 
Once all background processes will be predicted 
with the same accuracy, our predictions will make possible a realistic 
optimization of the~experimental~cuts.

This research was supported by
the HEPTOOLS EU program MRTN-CT-2006-035505,
the LHCPhenoNet network PITN-GA-2010-264564,
the Swiss National Science Foundation Joint Research Project SCOPES
IZ73Z0\_1/28079, the T\'AMOP 4.2.1./B-09/1/KONV-2010-0007 project,
the Hungarian Scientific Research Fund grant K-101482, 
the MEC project FPA 2008-02984 (FALCON). 
M.V.G and Z.T thank the Galileo Galilei Institute for Theoretical
Physics for the hospitality and the INFN partial support. We are grateful
to A. Tropiano, G. Dissertori, S. Moch and P. Skands for discussions. 

\end{document}